\documentclass[12pt,epsf]{article}
\usepackage{amsmath}
\usepackage[dvipdfmx]{graphicx}

\begin{document}

\newcommand{\bra}[1]{\left\langle #1 \right|}
\newcommand{\ket}[1]{\left| #1 \right\rangle}
\def \cJ{{ \cal J}}
\newcommand{\boldk}{\bm{k}}
\newcommand{\boldn}{\bm{n}}
\newcommand{\boldp}{\bm{p}}
\newcommand{\boldx}{\bm{x}}
\newcommand{\boldI}{\bm{I}}
\newcommand{\boldJ}{\bm{J}}
\newcommand{\boldr}{\bm{r}}
\newcommand{\boldR}{\bm{R}}
\newcommand{\boldtheta}{\bm{\theta}}
\newcommand{\boldomega}{\bm{\omega}}
\newcommand{\boldalpha}{\bm{\alpha}}
\newcommand{\boldpi}{\bm{\pi}}
\def\<{\langle}
\def\>{\rangle}
\def \beq{\begin{equation}}
\def \eeq{\end{equation}}
\def \ben{\begin{eqnarray}}
\def \een{\end{eqnarray}}
\def \bea{\begin{array}}
\def \eea{\end{array}}
\def \bem{\begin{displaymath}}  
\def \eem{\end{displaymath}}
\def \frac#1#2{ { #1 \over #2} }
\def \braketa#1{\langle #1 \rangle}
\def \braketb#1#2{\langle #1 | #2 \rangle}
\def \braketc#1#2#3{\langle #1 | #2 | #3 \rangle}
\def \gtsim{\mathrel{\hbox{\raise0.2ex
     \hbox{$>$}\kern-0.75em\raise-0.9ex\hbox{$\sim$}}}}
\def \ltsim{\mathrel{\hbox{\raise0.2ex
     \hbox{$<$}\kern-0.75em\raise-0.9ex\hbox{$\sim$}}}}
\def \dE{{ \mit\Delta E}}
\def \dI{{ \mit\Delta I}}
\def \cJ{{ \cal J}}
\def \cH{{ \cal H}}
\def \cQ{{ \cal Q}}
\def \cB{{ \cal B}}
\def \cF{{ \cal F}}
\def \cG{{ \cal G}}
\def \cW{{ \cal W}}
\def \cU{{ \cal U}}
\def \data{{ \rm data}}
\def \Nils{{ \rm Nils}}
\def \self{{ \rm self}}
\def \rexp{{ \rm exp}}
\def \rcal{{ \rm cal}}
\def \rms{{ \rm rms}}
\def \osc{{ \rm osc}}
\def \rot{{ \rm rot}}
\def \RPA{{ \rm RPA}}
\def \max{{ \rm max}}
\def \intr{{ \rm intr}}
\def \evod{{ {\rm e}\hbox{\raise-0.15ex\hbox{\rm -}}{\rm o} }}
\def \oneqp{{ 1\hbox{\raise-0.15ex\hbox{\rm -}}{\rm q.p.} }}
\def \nmax{{ n_{\rm max}}}
\def \colon{{ \mbox{:} }}
\def \maru#1{ {\mathop{#1}^{\circ}} }
\def \T{{ \rm T}}
\newcommand{\idp}{\int \frac{{\rm d}^3\bp}{(2\pi)^3}}
\newcommand{\idk}{\int \frac{{\rm d}^3\bk}{(2\pi)^3}}
\newcommand{\ip}{\int \frac{{\rm d}^4p}{(2\pi)^4}}
\newcommand{\ik}{\int \frac{{\rm d}^4k}{(2\pi)^4}}
\newcommand\bx{{\bf x}}
\newcommand\nn{\nonumber \\}
\renewcommand{\d}{\partial}
\newcommand{\dbeta}{\frac{\partial}{\partial\beta}}
\newcommand{\dgamma}{\frac{\partial}{\partial\gamma}}
\newcommand{\ddbeta}{\frac{{\partial}^2}{\partial{\beta}^2}}
\newcommand{\ddgamma}{\frac{{\partial}^2}{\partial{\gamma}^2}}
\newcommand{\dss}[1]{#1 \hspace{-0.45em}/} 
\newcommand{\ds}[1]{#1 \hspace{-0.5em}/}  
\newcommand{\Ds}[1]{#1 \hspace{-0.55em}/} 
\newcommand{\Dds}[1]{#1 \hspace{-0.6em}/} 
\newcommand\bzeta{\mbox{\boldmath$\zeta$}}
\newcommand\bgamma{\mbox{\boldmath$\gamma$}}
\newcommand\bsigma{\mbox{\boldmath$\sigma$}}
\newcommand\bSigma{\mbox{\boldmath$\Sigma$}}
\newcommand\btau{\mbox{\boldmath$\kappa$}}
\newcommand\btheta{\mbox{\boldmath$\theta$}}
\newcommand\bk{{\bf k}}
\newcommand\bq{{\bf q}}
\newcommand\E{\epsilon}
\newcommand\brho{\mbox{\boldmath$\rho$}}
\newcommand\bp{{\bf p}}
\def\rvec{\vec{r}}
\def\xvec{\vec{x}}
\def\half{{1\over 2}}
\def\sla{\slash{\!\!\!} }
\newcommand{\vp}{\mbox{\boldmath $p$}}
\newcommand{\vq}{\mbox{\boldmath $q$}}
\newcommand{\vr}{\mbox{\boldmath $r$}}
\newcommand{\vk}{\mbox{\boldmath $k$}}
\newcommand{\vP}{\mbox{\boldmath $P$}}
\newcommand{\vR}{\mbox{\boldmath $R$}}
\newcommand{\tpsp}{\hspace{1.5em}}
\newcommand{\del}{\partial}
\newcommand{\Hhat}{\hat{H}}
\newcommand{\Nhat}{\hat{N}}
\newcommand{\Qhat}{\hat{Q}}
\newcommand{\Phat}{\hat{P}}
\newcommand{\Ghat}{\hat{G}}
\newcommand{\That}{\hat{\Theta}}
\newcommand{\Nt}{\tilde{N}}
\newcommand{\Hc}{{\cal H}}
\newcommand{\bg}{\beta,\gamma}
\newcommand{\Dhatp}{\hat{D}^{(+)}}
\newcommand{\Dp}{D^{(+)}}
\newcommand{\Ddotp}{\dot{D}^{(+)}}
\newcommand{\adag}{a^{\dagger}}
\begin{center}
{\bf\Large 
Microscopic derivation of the quadrupole collective Hamiltonian 
for shape coexistence/mixing dynamics
}\\
\end{center}

\vspace{5mm}

\begin{center}
Kenichi Matsuyanagi,$^{1,2}$
Masayuki Matsuo,$^{3}$ 
Takashi Nakatsukasa,$^{1,4}$
\\
Kenichi Yoshida,$^{5}$
Nobuo Hinohara,$^{4,6}$ 
and Koichi Sato$^{1}$
\end{center}

\begin{center}
$^{1}$ RIKEN Nishina Center, Wako 351-0198, Japan\\
$^{2}$ Yukawa Institute for Theoretical Physics, Kyoto University, Kyoto 606-8502, Japan \\
$^{3}$ Department of Physics, Faculty of Science, Niigata University, Niigata 950-2181, Japan \\
$^{4}$ Center for Computational Sciences, University of Tsukuba, Tsukuba 305-8571, Japan \\
$^{5}$ Graduate School of Science and Technology, Niigata University, Niigata 950-2181, Japan \\
$^{6}$ National Superconducting Cyclotron Laboratory, Michigan State University, \\ 
          East Lansing, MI 48824-1321, USA
\end{center}

\abstract{
Assuming that the time-evolution of the self-consistent mean field 
is determined by five pairs of collective coordinate and collective momentum, 
we microscopically derive the collective Hamiltonian 
for low-frequency quadrupole modes of excitation. 
We show that the five-dimensional collective Schr\"odinger equation is capable of 
describing large-amplitude quadrupole shape dynamics 
seen as shape coexistence/mixing phenomena.    
We focus on basic ideas and recent advances of the approaches based 
on the time-dependent mean-field theory, but relations to other 
time-independent approaches are also briefly discussed. 
}

\section{Introduction}

In this paper, we focus on low-frequency quadrupole motions 
which play the major role in low-energy excitation spectra.
As is well known, various giant resonances appearing in highly excited states are 
well described by the random-phase approximation (RPA),  
which is a small-amplitude approximation of 
the time-dependent Hartree-Fock (TDHF) theory. 
In contrast to the giant resonances, low-frequency quadrupole vibrations 
exhibit characteristic features associated with superfluidity of  
the finite quantum system (nucleus), that is, 
pairing correlations and varying shell structure of the
self-consistent mean field play essential roles 
\cite{boh75,abe90,ben03,row10}. 
\\

\noindent
{\it Quantum shape fluctuation, shape mixing and shape coexistence}\\

The low-frequency quadrupole vibrations can be regarded as 
soft modes of the quantum phase transition towards 
equilibrium deformations of the mean field. 
As is well known, in nuclei situated in the transitional region from spherical to deformed, 
amplitudes of quantum shape fluctuation about the equilibrium remarkably increase. 
This is the case also for weakly deformed nuclei where the gain in binding energies
due to the symmetry breaking is comparable in magnitude to the vibrational zero-point energies. 
The transitional region is wide and those nuclei exhibit quite rich excitation spectra.  
This is a characteristic feature of finite quantum systems and provides an invaluable opportunity 
to investigate the process of the quantum phase transition through analysis of quantum spectra.
To describe such large-amplitude collective motions (LACM),  
we need to go beyond the small-amplitude approximation (quasiparticle RPA) of 
the time-dependent Hartree-Fock-Bogoliubov (TDHFB) theory for superfluid systems. 
It is required to develop a microscopic theory of LACM
capable of describing the varying shell structure associated with the 
time-dependent mean field with superfluidity.   

The spherical shell structure gradually changes with the growth of deformation 
and generates `deformed shell structures' and `deformed magic numbers,' 
that stabilize certain deformed shapes of the mean field. 
When a few local minima of the mean field with different shapes appear in the same energy region, 
LACM tunneling through potential barriers and extending between local minima may take place. 
These phenomena may be regarded as a kind of macroscopic quantum tunneling. 
Note that the barriers are not generated by external fields but self-consistently generated as 
a consequence of quantum dynamics of the many-body system under consideration. 
Quantum spectra of low-energy excitation that needs such concepts have been observed 
in almost all regions of the nuclear chart 
\cite{hey11}. 
When different kinds of quantum eigenstates associated with different shapes coexist 
in the same energy region, 
we may call them `{\it shape coexistence phenomena.}'
This is the case when shape mixing due to tunneling motion is weak and 
collective wave functions retain their localizations about different equilibrium shapes. 
On the other hand, if the shape mixing is strong, 
large-amplitude shape fluctuations ({\it delocalization} of the collective wave functions) 
extending to different local minima may occur. 
\\

\noindent
{\it Collective rotations restoring broken symmetries}\\

As is well known, the central concept of the BCS theory of superconductivity is 
spontaneous breaking of the gauge symmetry and emergence of collective modes. 
The massless modes restoring the broken symmetry   
are called Anderson-Nambu-Goldstone (ANG) modes   
\cite{and58,nam60,bri05}. 
As emphasized by Bohr and Mottelson, nuclear rotation can be regarded as 
an ANG mode restoring the broken rotational symmetry in real space    
\cite{boh75,fra01}.

In finite quantum systems such as nuclei, the rotational ANG modes may couple 
rather strongly with quantum shape-fluctuation modes. 
For instance, even when the self-consistent mean field acquires  
a deep local minimum at a finite value of $\beta$, 
the nucleus may exhibit a large-amplitude shape fluctuation 
in the $\gamma$ degree of freedom, 
if the deformation potential is flat in this direction. 
Here, as usual, $\beta$ and $\gamma$ represent the magnitudes of 
axially symmetric and asymmetric quadrupole deformations, respectively.  
Such a situation is widely observed in experiments and called $\gamma$-soft. 
Although the quantum-mechanical collective rotation is forbidden about 
the symmetry axis, the rotational degrees of freedoms about three principal axes are 
all activated once the axial symmetry is dynamically broken due to the 
quantum shape fluctuation. Rotational spectra in such $\gamma$-soft nuclei 
do not exhibit a simple $I(I+1)$ pattern. 
Such an interplay of the shape-fluctuation and rotational modes may be regarded as 
a characteristic feature of finite quantum systems and provides an invaluable opportunity 
to investigate the process of the quantum phase transition through analysis of quantum spectra. 

Thus, we need to treat the two kinds of collective variables, {\it i.e.}, those associated with 
the symmetry-restoring ANG modes and those for quantum shape fluctuations, in a unified manner 
to describe low-energy excitation spectra of nuclei. 
\\

\noindent
{\it Five-dimensional quadrupole collective Hamiltonian}\\

Vibrational and rotational motions of the nucleus can be described as 
time-evolution of a self-consistent mean field. 
This is the basic idea underlying the unified model of Bohr and Mottelson 
\cite{boh76,mot76}.
In this approach, the five-dimensional (5D) collective Hamiltonian
describing the quadrupole vibrational and rotational motions is given by 
\cite{boh75,pro09}  
\begin{equation}
H_{\rm coll}=T_{\rm vib}+T_{\rm rot}+V(\beta,\gamma), 
\end{equation}
\begin{equation}
T_{\rm vib}=\frac{1}{2}D_{\beta\beta}(\beta,\gamma)\dot 
\beta^2+D_{\beta\gamma}(\beta,\gamma)\dot \beta \dot
\gamma+\frac{1}{2}D_{\gamma\gamma}(\beta,\gamma)\dot \gamma^2, 
\end{equation}
\begin{equation}
T_{\rm rot}=\sum_{k}\frac{I_k^2}{2\cJ_k(\beta,\gamma)}. 
\end{equation}
Here, $\beta$ and $\gamma$ are treated as dynamical variables, and 
$\dot \beta$ and $\dot \gamma$ represent their time-derivatives. 
They are related to expectation values of the quadruple operators 
(with respect to the time-dependent mean-field states) and their variations in time.    
The quantities ($D_{\beta\beta}, D_{\beta\gamma}$, and $D_{\gamma\gamma}$) 
appearing in the kinetic energies of vibrational motion, $T_{\rm vib}$, 
represent inertial masses of the vibrational motion. 
They are functions of $\beta$ and $\gamma$.  
The quantities $I_k$ and $\cJ_k (\beta,\gamma)$ 
in the rotational energy $T_{\rm rot}$
represent the three components of the angular momentum and 
the corresponding moments of inertia, respectively. 
Note that they are defined with respect to the principal axes of the 
body-fixed (intrinsic) frame that is attached to the instantaneous shape of 
the time-dependent mean-field. 

In the case that the potential energy $V(\beta,\gamma)$ 
has a deep minimum at a finite value of $\beta$ and 
$\gamma=0^\circ$ (or $\gamma=60^\circ$), 
a regular rotational spectrum with the $I(I+1)$ pattern may appear. 
In addition to the ground band,  we can expect the $\beta$ and $\gamma$ bands to appear, 
where vibrational quanta with respect to the $\beta$ and $\gamma$ degrees of freedom 
are excited. 
Detailed investigations on the $\gamma$-vibrational bands over many nuclei 
have revealed, however, that they usually exhibit significant anharmonicities (non-linearlities) 
\cite{mat85b}.
Also for the $\beta$-vibrational bands, it has been known \cite{iwa76, wee81, tak86} 
that they couple, sometimes very strongly, with pairing-vibrational modes 
(associated with fluctuations of the pairing gap). 
Recent experimental data indicate the need for a radical review of their characters 
\cite{hey11}. 
\\

\noindent
{\it Collective quantization of time-dependent mean fields}\\

States vectors of time-dependent mean field are kinds of generalized coherent states, 
and we can rigorously formulate the TDHFB as 
a theory of classical Hamiltonian dynamical system of large dimension 
\cite{neg82,abe83,yam87,kur01}.  
Because time-evolution of the mean field is determined by the classical Hamilton equations, 
we cannot describe, within the framework of the TDHFB, quantum spectra of low-lying states 
and macroscopic quantum tunneling phenomena such as spontaneous fissions and subbarrier fusions. 
To describe these genuine quantum phenomena, 
we need to introduce a few collective variables determining the time-evolution of the 
mean field and quantize them. We refer this procedure `{\it collective quantization.}' 

For small-amplitude vibrations about an HFB equilibrium, it is well known that 
we can introduce collective variables in a microscopic way 
by solving the quasiparticle RPA (QRPA) equations. 
Single-particle spectra for a mean-field of a finite quantum system have rich shell structures, 
and thereby a variety of collective vibrational modes emerge.  
Even within the isoscalar quadrupole vibrations, two collective modes appear exhibiting 
quite different characteristics; 
the low- (usually first excited $2^+$) and high-frequency (giant resonance) modes.  
One of the merits of the QRPA is that we can determine the microscopic structures of 
the collective coordinates and momenta starting from a huge number of 
microscopic (particle-hole, particle-particle, and hole-hole) degrees of freedom. 
We can thus learn how collective vibrations are generated as 
coherent superpositions of many two-quasiparticle excitations. 
Examining the microscopic structure of the low-frequency quadrupole vibrations, 
we see that the weights of two-quasiparticle excitations near the Fermi surface 
are much larger than those in the mass quadrupole operators 
(see, {\it e.g.}, Ref. \cite{nak99}).  
This clearly indicates the importance of describing collective modes in a microscopic way. 
Another merit of the QRPA is that it yields the ANG modes and 
their collective masses (inertial functions).  
In this way, we can restore the symmetries broken by the mean-field approximation 
\cite{bri05, rin80, bla86}.

It has been one of the longstanding fundamental subjects in nuclear structure physics 
to construct a microscopic theory of LACM  
by extending the QRPA concepts to non-equilibrium states 
\cite{mat10, nak12, mat13}. 
Below we shall briefly review various ideas proposed up to now for this aim.

\section{Basic ideas of large-amplitude collective motion}

During the attempts to construct a microscopic theory of LACM  
since the latter half of the 1970s, significant progress has been achieved in the 
fundamental concepts of collective motion. Especially important is the recognition that
microscopic derivation of the collective Hamiltonian is equivalent to extraction of a collective
submanifold embedded in the TDHFB phase space, which is approximately decoupled
from other non-collective degrees of freedom. From this point of view we can say that 
collective variables are nothing but local canonical variables which can be flexibly chosen
on this submanifold. Below we review recent developments achieved on the basis of such
concepts.
\\

\noindent
{\it Extraction of collective submanifold}\\

Attempts to formulate a LACM theory 
without assuming adiabaticity of large-amplitude collective motion 
were initiated 
by Rowe and Bassermann \cite{row76} and Marumori \cite{mar77},  
and led to the formulation of the self-consistent collective coordinate (SCC) method 
\cite{mar80}. 
In these approaches, collective coordinates and collective momenta are treated on the same footing.  
In the SCC method, basic equations determining the collective submanifold are derived 
by requiring maximal decoupling of the collective motion of interest and 
other non-collective degrees of freedom. 
The collective submanifold is a geometrical object that is invariant with respect to the choice of 
the coordinate system whereas the collective coordinates depend on it.  
The idea of coordinate-independent theory of collective motion was developed also   
by Rowe \cite{row82} and Yamamura and Kuriyama \cite{yam87}. 
This idea gave a deep impact on the fundamental question `{\it what are the collective variables.}' 
The SCC method was first formulated for TDHF, but 
later extended to TDHFB for describing nuclei with superfluidity 
\cite{mat86}. 

In the SCC method, under the assumption that time evolution is governed by 
a few collective coordinates $q=(q_1,q_2,...,q_f)$  and collective momenta $p=(p_1,p_2,...,p_f)$, 
the TDHFB states vectors are written as 
\begin{equation}
\ket{\phi(t)}  = \ket{\phi(q,p)} = e^{i{\hat G}(q, p)} \ket{\phi_0}, 
\end{equation}
or equivalently, 
\begin{equation}
\ket{\phi(t)} = \ket{\phi(\eta,\eta^*)} = e^{i{\hat G}(\eta, \eta^*)} \ket{\phi_0}, 
\end{equation}
where $\ket{\phi_0}$ denotes the HFB ground state and $\eta=(\eta_1,\eta_2,...,\eta_f)$ with  
\begin{equation}
\eta_i = \frac{1}{\sqrt 2}(q_i+ip_i),~~~~~\eta_i^* = \frac{1}{\sqrt 2}(q_i-ip_i).
\end{equation}

The TDHFB states $\ket{\phi(t)}$ are required to fulfill the canonical variable conditions 
\begin{eqnarray}
\bra{\phi(\eta,\eta^*)} \frac{\del}{\del\eta_i} \ket{\phi(\eta,\eta^*)} &=& \frac{1}{2}\eta_i^*, \\
\bra{\phi(\eta,\eta^*)} \frac{\del}{\del\eta_i^*} \ket{\phi(\eta,\eta^*)} &=& -\frac{1}{2}\eta_i, 
\end{eqnarray}
which guarantee that $(q,p)$ are canonical conjugate pairs. 
The one-body operator ${\hat G}(\eta, \eta^*)$ is determined by  
the time-dependent variational principle 
\begin{equation}
\delta\bra{\phi(\eta,\eta^*)} i\frac{\partial}{\partial t} - H \ket{\phi(\eta,\eta^*)}=0  
\end{equation}
with  
\begin{equation}
\frac{\partial}{\partial t} = \sum_i 
(\dot{\eta_i}\frac{\partial}{\partial \eta_i} + 
\dot{\eta_i}^*\frac{\partial}{\partial \eta_i^*} ), 
\end{equation}
and the canonical variable conditions. 

Making a power-series expansion of ${\hat G}$ with respect to $(\eta,\eta^*)$,
\begin{equation}
{\hat G}(\eta,\eta^*) = \sum_{ij} \sum_{m_i n_j} {\hat G}_{m_i n_j} (\eta_i^*)^{m_i} \eta_j^{n_j},           
\end{equation}
and requiring that the time-dependent variational principle holds for every power of $(\eta,\eta^*)$, 
we can successively determine the one-body operator ${\hat G}_{m_i n_j}$. 
This method of solution is called the $(\eta,\eta^*)$ expansion method. 
The collective Hamiltonian is defined by the expectation value of the microscopic Hamiltonian 
with respect to $\ket{\phi(\eta,\eta^*)}$. Because $(\eta,\eta^*)$ are canonical variables, 
they are replaced by boson operators after canonical quantization. 
The lowest linear order corresponds to the QRPA.  
Accordingly, the collective variables $(\eta_i,\eta_i^*)$ correspond to specific QRPA modes  
in the small-amplitude limit. It is important to note, however, that 
the microscopic structure of ${\hat G}$ changes as a function of $(\eta_i,\eta_i^*)$ 
due to the mode-mode coupling effects among different QRPA modes in the higher order. 
In this sense, the SCC method may be regarded 
as a dynamical extension of the boson expansion method 
\cite{mat85a}.  
Thus, the SCC is a powerful method of treating anharmonic effects to the QRPA vibrations 
originating from mode-mode couplings, as shown in its application to the two-phonon states 
of anharmonic $\gamma$ vibration 
\cite{mat85b,mat85c}. 
The SCC method was also used for derivation of the 5D quadrupole collective Hamiltonian and analysis of 
the quantum phase transition from spherical to deformed shapes 
\cite{yam93}, 
and for constructing diabatic representation in the rotating shell model 
\cite{shi01}. 
\\

\noindent
{\it Solution with adiabatic expansion }\\

The $(\eta,\eta^*)$ expansion about an HFB equilibrium point is not suitable for treating 
situations such as shape coexistence, 
where a few local minima energetically compete in the HFB potential energy surface. 
For describing adiabatic LACM extending over different HFB local minima, 
a new method has been proposed 
\cite{mat00}. 
In this method, 
the basic equations of the SCC method are solved by an expansion with respect to 
the collective momenta. It is called `adiabatic SCC (ASCC) method.'  
Similar methods have been developed also 
by Klein {\it et al.} \cite{kle91a}, and Almehed and Walet \cite{alm04a}.

Let us assume that the TDHFB state vector can be written as 
\begin{equation}
 \ket{\phi(q,p)}  =  e^{i \sum_i p_i \Qhat^i(q) } \ket{\phi(q)}. 
\label{eq:ASCCstate}
\end{equation}
Here, $\Qhat^i(q)$ are one-body operators, called infinitesimal generators,  
and $\ket{\phi(q)}$ is an intrinsic state at the collective coordinate $q$, 
called a moving frame HFB state. 

We determine the microscopic structures of $\Qhat^i(q)$ and $\ket{\phi(q)}$    
by the time-dependent variational principle  
\begin{equation}
\delta \bra{\phi(q,p)} i\frac{\del}{\del t} - \Hhat \ket{\phi(q,p)}  = 0.
\label{eq:TDVP}
\end{equation}

Making power-series expansions with respect to the collective momenta $p$ 
and retaining terms up to the second order in $p$,  
we obtain\\

\underline{moving-frame HFB equation}
\begin{equation}
 \delta\bra{\phi(q)}\Hhat_M(q)\ket{\phi(q)} = 0, 
\label{eq:mfHFB}
\end{equation}

\underline{moving-frame QRPA equations (local harmonic equations)}

\begin{eqnarray}
 \delta\bra{\phi(q)} [\Hhat_M(q), \Qhat^i(q)]
- \frac{1}{i} \sum_{k} B^{ik}(q) \Phat_k(q)
+ \frac{1}{2}\left[\sum_{k} \frac{\del V}{\del q^k}\Qhat^k(q),
 \Qhat^i(q)\right]
\ket{\phi(q)} = 0, 
\label{eq:ASCC1}
\end{eqnarray}
\begin{eqnarray}
 \delta\bra{\phi(q)} [\Hhat_M(q),
 \frac{1}{i}\Phat_i(q)] - \sum_{j} C_{ij}(q) \Qhat^j(q) 
 - \frac{1}{2}\left[\left[\Hhat_M(q), \sum_{k} \frac{\del V}{\del
 q^k}\Qhat^k(q)\right], \sum_{j} B_{ij}(q) \Qhat^j(q)\right]
\nonumber \\  
\ket{\phi(q)}
 = 0, 
\label{eq:ASCC2}
\end{eqnarray}
where $\Hhat_M(q)$ represents the Hamiltonian in the frame attached to the moving mean field, 
\begin{eqnarray}
  \Hhat_M(q) = 
 \Hhat 
- \sum_{i} \frac{\del V}{\del q^i}\Qhat^i(q), 
\end{eqnarray}
and is called `moving-frame Hamiltonian.' 
The displacement operators $\Phat_i(q)$ and $C_{ij}(q)$ are defined by 
\begin{eqnarray}
 \Phat_i(q) \ket{\phi(q)} = i \frac{\del}{\del q^i} \ket{\phi(q)} 
\end{eqnarray}
and
\begin{eqnarray}
 C_{ij}(q) = \frac{\del^2 V}{\del q^i \del q^j} - \sum_k \Gamma_{ij}^k\frac{\del V}{\del q^k}, 
\end{eqnarray}
respectively, with
\begin{eqnarray}
 \Gamma_{ij}^k(q) = \frac{1}{2} \sum_l B^{kl}( \frac{\del B_{li}}{\del q^j}
 + \frac{\del B_{lj}}{\del q^i} - \frac{\del B_{ij}}{\del q^l} ). 
\end{eqnarray}
The double-commutator term in Eq.~(\ref{eq:ASCC2}) arises from the $q$-derivative 
of the infinitesimal generators $\Qhat^i(q)$ and 
represents the curvatures of the collective submanifold.  

Solving these equations self-consistently, we can determine the microscopic expressions of 
the infinitesimal generators $\Qhat^i(q)$ and $\Phat_i(q)$ 
in bilinear forms of the quasiparticle creation and annihilation operators 
defined locally with respect to $\ket{\phi(q)}$.
The collective Hamiltonian is given by 
\begin{eqnarray}
 \Hc(q,p) 
&=&\bra{\phi(q,p)}\Hhat
 \ket{\phi(q,p)} \nonumber \\
&=&V(q) + \frac{1}{2} \sum_{ij} B^{ij}(q)p_i p_j 
\label{eq:collH_ASCC}
\end{eqnarray}
with 
\begin{eqnarray}
 V(q) =  \Hc(q,p)
\Big\arrowvert_{p=0}, ~~~~~
 B^{ij}(q) = \frac{\del^2 \Hc}{\del p_i \del p_j}
 \Big\arrowvert_{p=0}.  \label{eq:defB}   
\label{eq:deflambda}
\end{eqnarray}
where $V(q)$ and $B^{ij}(q)$ represent the collective potential and  
the reciprocals of collective inertial mass, respectively. 
They are functions of the collective coordinate $q$. 
Note that Eqs.~(\ref{eq:mfHFB}), (\ref{eq:ASCC1}), and (\ref{eq:ASCC2}) 
reduce to the HFB and QRPA equations at equilibrium points, where 
$\del V/\del q^i=0$; namely, they are natural extensions of the HFB-QRPA equations 
to non-equilibrium states. 

Let us note the following points.  
\begin{itemize}
\item {\it Difference from the constrained HFB equations} \\
The moving-frame HFB equation (\ref{eq:mfHFB}) resembles 
the constrained HFB equation, but the infinitesimal generators $\Qhat^i(q)$ 
are here self-consistently determined together with $\Phat_i(q)$ as solutions of the 
moving-frame QRPA equations (\ref{eq:ASCC1}) and (\ref{eq:ASCC2}) 
at every point of the collective coordinate $q$. 
Thus, contrary to constrained operators in the constrained HFB theory, 
their microscopic structures change as functions of $q$. 
In other words, the optimal `constraining' operators are locally determined at every point of $q$. 
The collective submanifold embedded in the TDHFB phase space is 
extracted in this way.  The canonical quantization of the collective Hamiltonian 
described by a few collective variables $(q,p)$ is similar to the quantization 
of constrained system 
\cite{row82}, 
but the `constraints' are here  
generated by the dynamics of the quantum many-body system under consideration. 
\item {\it Meaning of the term `adiabatic'} \\
It is used here in the meaning that we can solve 
the time-dependent variational equation (\ref{eq:TDVP}) in a good approximation 
by taking into account up to the second order 
in an expansion with respect to the collective momenta $p$.  
It is important to note that the effects of finite frequency of the LACM are taken into account 
through the moving-frame QRPA equations. No assumption is made like that 
the kinetic energy of LACM is much smaller 
than the lowest two-quasiparticle excitation energy at every point of $q$.  
\item {\it Physics of collective inertial mass} \\
The collective inertial mass represents the inertia of the many-body system 
against an infinitesimal change of the collective coordinate $q$
during the time evolution of the mean field. 
It is a local quantity and varies as a function of $q$. 
As the single-particle-energy spectrum in the mean field changes 
as a function of $q$, level crossing at the Fermi energy successively occurs 
during the LACM.  In the presence of the pairing correlation,  
the many-body system can easily rearrange the lowest-energy  
configurations at every value of $q$, {\it i.e.}, the system can easily change $q$.    
The easiness/hardness of the configuration rearrangements at the 
level crossings determines the adiabaticity/diabaticity of the system. 
Since the inertia represents a property of the system trying 
to keep a definite configuration, we expect that  
the stronger the pairing correlation becomes, 
the smaller the collective inertial mass becomes  
\cite{bar90}. 
It remains as an interesting subject to investigate how the self-consistent determination of 
the optimal directions of collective motion and the finite frequency $\omega(q)$ 
of the moving-frame QRPA modes 
affect the level-crossing dynamics of the superfluid nuclear systems. 
\end{itemize}

\noindent
{\it Consideration of gauge invariance}\\
  
In the QRPA, the ANG modes such as the number-fluctuation (pairing rotational) modes 
are decoupled from other normal modes 
and thereby the QRPA restores the gauge invariance (number conservation) 
broken in the HFB mean field 
\cite{bri05}. 
It is desirable to keep such a merits of the QRPA beyond the small-amplitude approximation.   
Otherwise, spurious number-fluctuation modes would unexpectedly mix in the LACM of interest. 
We can take into account the gauge invariance in the following way 
\cite{hin07}. 
Introducing the number-fluctuation variables $n^{(\tau)}$ and 
the gauge angles $\varphi^{(\tau)}$ conjugate to them,  
we write the TDHFB state vector (\ref{eq:ASCCstate}) in a more general form: 
\begin{equation}
 \ket{\phi(q,p,\varphi,n)} =
 e^{-i\sum_{\tau} \varphi^{(\tau)}\Nt^{(\tau)}} \ket{\phi(q,p,n)} 
 \end{equation}
 with
 \begin{equation}
 \ket{\phi(q,p,n)}  = e^{i \Ghat(q,p,n)} \ket{\phi(q)} 
 \end{equation} 
 and
 \begin{equation}
 \Ghat(q,p,n) = \sum_i p_i \Qhat^i(q) + \sum_{\tau=n,p} n^{(\tau)} \That^{(\tau)}(q). 
\end{equation}
Here 
$\Nt^{(\tau)}\equiv\Nhat^{(\tau)}- N_0^{(\tau)}$ 
are the number-fluctuation operators 
measured from $N_0^{(\tau)}\equiv\bra{\phi(q)}\Nhat^{(\tau)}\ket{\phi(q)}$,  
with the suffix $\tau$ distinguishing protons and neutrons, 
and $\That^{(\tau)}(q)$ infinitesimal generators for the pairing-rotation degrees of freedom.
The state vector $\ket{\phi(q,p,n)}$ may be regarded as an intrinsic state to the pairing rotation.  
Using $\ket{\phi(q,p,\varphi,n)}$ in place of $\ket{\phi(q,p)}$ in Eq.~(\ref{eq:TDVP}) 
and expanding it in $n^{(\tau)}$ as well as $p$ up to the second order, 
we can determine $\That^{(\tau)}(q)$ simultaneously with $\Qhat^i(q)$ and $\Phat^i(q)$ 
such that the moving-frame HFB+QRPA equations become invariant 
against the rotation of the gauge angle $\varphi^{(\tau)}$. 
The gauge invariance of the resulting equations implies that we need 
to fix a gauge in numerical applications. A convenient  procedure of the gauge fixing is discussed in 
Ref. \cite{hin07}. 

\section{Microscopic derivation of the 5D quadrupole collective Hamiltonian}

For collective submanifolds of two dimensions (2D) or more, large-scale numerical computation is needed 
to find fully self-consistent solutions of the ASCC equations.  
Then, a practical approximation scheme, called local QRPA (LQRPA) method, has been 
developed 
\cite{hin10,sat11,sat12}. 
This scheme may be regarded as a first step of iterative solution of Eqs.~(\ref{eq:mfHFB}) - (\ref{eq:ASCC2}). 
With use of it, we can easily derive the 5D collective Hamiltonian.  
We first derive the 2D collective Hamiltonian for vibrational motions corresponding to 
the $(\beta,\gamma)$ deformations, and then consider the three-dimensional (3D) rotational motions. 

First, we solve the constrained HFB equation 
\begin{eqnarray}
 \delta \bra{\phi(\bg)} \Hhat_{\rm CHFB}(\bg) \ket{\phi(\bg)} = 0      
\label{eq:CHFB} 
\end{eqnarray}
with
\begin{eqnarray}
 \Hhat_{\rm CHFB}(\bg) = \Hhat -
 \sum_{\tau}\lambda^{(\tau)}(\bg)\Nt^{(\tau)} \nonumber 
  - \sum_{m = 0, 2} \mu_{m}(\bg) \Dhatp_{2m}, 
\label{eq:H_CHFB} 
\end{eqnarray}
where $\lambda^{(\tau)}(\bg)$, $\mu_{m}(\bg)$ and $\Dhatp_{2m}$ 
are the chemical potentials, the Lagrange multipliers,  
and the quadrupole operators, respectively. 
The quadrupole deformation parameters $(\beta,\gamma)$ are defined by 
\begin{eqnarray}
 \beta\cos\gamma &=  \eta D^{(+)}_{20} 
= \eta \bra{\phi(\bg)} \Dhatp_{20} \ket{\phi(\bg)},  \label{eq:definition1}  \\
 \frac{1}{\sqrt{2}} \beta\sin\gamma &= \eta D^{(+)}_{22} 
= \eta \bra{\phi(\bg)} \Dhatp_{22} \ket{\phi(\bg)}, 
\label{eq:definition2} 
\end{eqnarray}
where $\eta$ is a scaling factor \cite{boh75,bar65}. 

Next, we solve 
\begin{eqnarray}
 \delta \bra{\phi(\bg)} [ \Hhat_{\rm CHFB}(\bg), \Qhat^i(\bg) ] 
 - \frac{1}{i} \Phat_i(\bg) \ket{\phi(\bg)}  = 0, \quad (i=1, 2) \\
\label{eq:LQRPA1}
 \delta \bra{\phi(\bg)} [ \Hhat_{\rm CHFB}(\bg), \frac{1}{i} \Phat_i(\bg)]
 - C_i(\bg) \Qhat^i(\bg) \ket{\phi(\bg)} = 0.  \quad (i=1, 2) 
\label{eq:LQRPA2}
\end{eqnarray}
These are the moving-frame QRPA equations without the curvature terms and 
called local QRPA (LQRPA) equations. 
Making a similarity transformation such that the collective masses 
corresponding to the collective coordinates $(q_1,q_2)$ become unity,  
we can write the kinetic energy of vibrational motions as   
\begin{eqnarray}
 T_{\rm vib} = \frac{1}{2} \sum_{i=1,2} (p_i)^2 
 = \frac{1}{2} \sum_{i=1,2} (\dot{q}^i)^2 
\label{eq:Tvib}
\end{eqnarray}  
without loss of generality. 
Changes of the quadrupole deformation due to variations 
with respect to $(q_1,q_2)$ are given by 
\begin{eqnarray}
d\Dp_{2m}   = \sum_{i=1,2} \frac{\del \Dp_{2m}}{\del q^i} dq^i.  \quad\quad (m=0,2) 
\end{eqnarray}

Thus, the kinetic energy of vibrational motions is given in terms of 
time derivatives of the quadrupole deformation,  
\begin{eqnarray}
 T_{\rm vib} = \frac{1}{2} M_{00} [\Ddotp_{20}]^2
  + M_{02} \Ddotp_{20} \Ddotp_{22}+ \frac{1}{2} M_{22} [\Ddotp_{22}]^2,
\label{eq:TvibD}
\end{eqnarray}
where 
\begin{eqnarray}
 M_{mm'}(\bg) = \sum_{i=1,2} \frac{\del q^i}{\del
 \Dp_{2m}} \frac{\del q^i}{\del \Dp_{2m'}}. 
\label{eq:M_mm}
\end{eqnarray}
It is straightforward to rewrite the above expression 
using the time derivatives of $(\beta,\gamma)$. 
Subsequently we solve the LQRPA equations for 3D rotational motions at every point 
on the $(\beta,\gamma)$ plane to obtain the inertial functions $D_k(\beta,\gamma)$ and 
the moments of inertia 
$\cJ_k(\beta,\gamma)=4\beta^2D_k(\beta,\gamma) \sin^2(\gamma-2\pi k/3)$ 
determining the rotational energy $T_{\rm rot}$. 
This step is the same as in Thouless and Valatin \cite{tho62}, 
except that the procedure is applied for non-equilibrium points of $(\beta,\gamma)$ 
as well as the equilibrium points in the potential energy surface $V(\beta, \gamma)$. 

After quantizing in curvilinear coordinates (so-called Pauli prescription)  \cite{eis87}, 
we obtain the quantized 5D quadrupole collective Hamiltonian, 
\begin{equation}
{\hat H}_{\rm coll}={\hat T}_{\rm vib}+{\hat T}_{\rm rot}+V(\beta,\gamma), 
\end{equation}
whose vibrational kinetic-energy term takes the following form: 
\begin{eqnarray}
{\hat T}_{\rm vib}=\frac{-\hbar^2}{2\sqrt{WR}}\left\{ \frac{1}{\beta^3} 
\left[\dbeta \left(\beta^3\sqrt{\frac{R}{W}}D_{\gamma\gamma}
\dbeta\right)-\dbeta \left(\beta^3\sqrt{\frac{R}{W}}D_{\beta\gamma}\dgamma\right)\right] 
\right. \\ \left.
+\frac{1}{\sin 3\gamma}\left[-\dgamma 
\left(\sqrt{\frac{R}{W}}\sin 3\gamma D_{\beta\gamma}\dbeta\right)
+\dgamma \left(\sqrt{\frac{R}{W}}\sin 3\gamma D_{\beta\beta}\dgamma\right)
\right] 
\right\}, 
\end{eqnarray}
where 
\begin{eqnarray}
W&=&D_{\beta\beta}(\beta,\gamma)D_{\gamma\gamma}(\beta,\gamma)-
D_{\beta\gamma}^2(\beta,\gamma), \\
R&=&D_1(\beta,\gamma)D_2(\beta,\gamma)D_3(\beta,\gamma). 
\end{eqnarray} 
If the inertial functions 
$(D_{\beta\beta}, D_{\gamma\gamma}, D_1, D_2, D_3)$ 
are replaced by a common constant $D$ and $D_{\beta\gamma}$ by 0, 
the above expression is reduced to 
\begin{equation}
{\hat T}_{\rm vib}=-\frac{\hbar^2}{2D} \left( {\frac{1}{\beta^4}} 
\dbeta \beta^4 \dbeta 
+{\frac{1}{\beta^2 \sin 3\gamma}}\dgamma \sin 3\gamma \dgamma \right). 
\end{equation}
Such a drastic approximation may be valid only for small-amplitude vibrations about a spherical HFB equilibrium. 

Writing the collective wave functions as 
\begin{eqnarray}
 \Psi_{\alpha IM}(\beta,\gamma,\Omega) &=& \sum_{K={\rm even}}\Phi_{\alpha IK}(\beta,\gamma) 
 \langle\Omega|IMK\rangle, \\
 \langle\Omega|IMK\rangle &=&
 \sqrt{\frac{2I+1}{16\pi^2(1+\delta_{K0})}}
 \left[ {\cal D}^I_{MK}(\Omega) + (-)^I {\cal D}^I_{M-K}(\Omega) \right]
\end{eqnarray}
and solving the eigenvalue equation for vibrational wave functions 
\begin{eqnarray}
 \left[ \hat{T}_{\rm vib} + V(\beta,\gamma) \right] \Phi_{\alpha IK}(\beta,\gamma)
 + \sum_{K'={\rm even}}
 \bra{IMK}\hat{T}_{\rm rot}\ket{IMK'} \Phi_{\alpha IK'}(\beta,\gamma)  = E_{\alpha I}
 \Phi_{\alpha IK}(\beta,\gamma),
\end{eqnarray}
we obtain quantum spectra of quadrupole collective motion. 
The symmetries and boundary conditions of the vibrational wave functions are 
discussed in Ref. \cite{kum67}. 

\section{Illustrative examples}

We here present some applications of the LQRPA method 
for deriving the 5D collective Hamiltonian. 
In the numerical examples below,  
the pairing-plus-quadrupole (P+Q) model Hamiltonian \cite{bes69} 
(including the quadrupole-pairing interaction) 
is employed in solving the CHFB + LQRPA equations. 
The single-particle energies and the P+Q interaction strengths 
are determined such that the results of the Skyrme-HFB calculation  
for the ground states are best reproduced within the P+Q model 
(see Refs.~\cite{sat11, hin11a}  for details). 
The LQRPA method is quite general 
and it can be used in conjunction with various Skyrme interactions 
or modern density functionals. 
A large-scale calculation is required, however, and 
such an application of the LQRPA method with realistic interactions/functionals 
is a challenging future subject.
A step toward this goal has recently been carried out 
for axially symmetric cases \cite{yos11}. 
More examples can be found in 
\cite{hin09} for $^{68-72}$Se, 
\cite{sat11} for $^{72,74,76}$Kr, 
\cite{hin10} for the $^{26}$Mg region,  
\cite{hin11b} for $^{30-34}$Mg, 
\cite{yos11} for $^{58-68}$Cr, 
\cite{sat12} for $^{58-66}$Cr, 
and
\cite{hin12} for $^{128-132}$Xe and $^{130-134}$Ba.  
\\

\noindent
{\it Oblate-prolate shape coexistence and fluctuations in $^{72}$Kr} 
\\

\begin{figure}[tbp]
\begin{center}
\includegraphics[width=\textwidth]{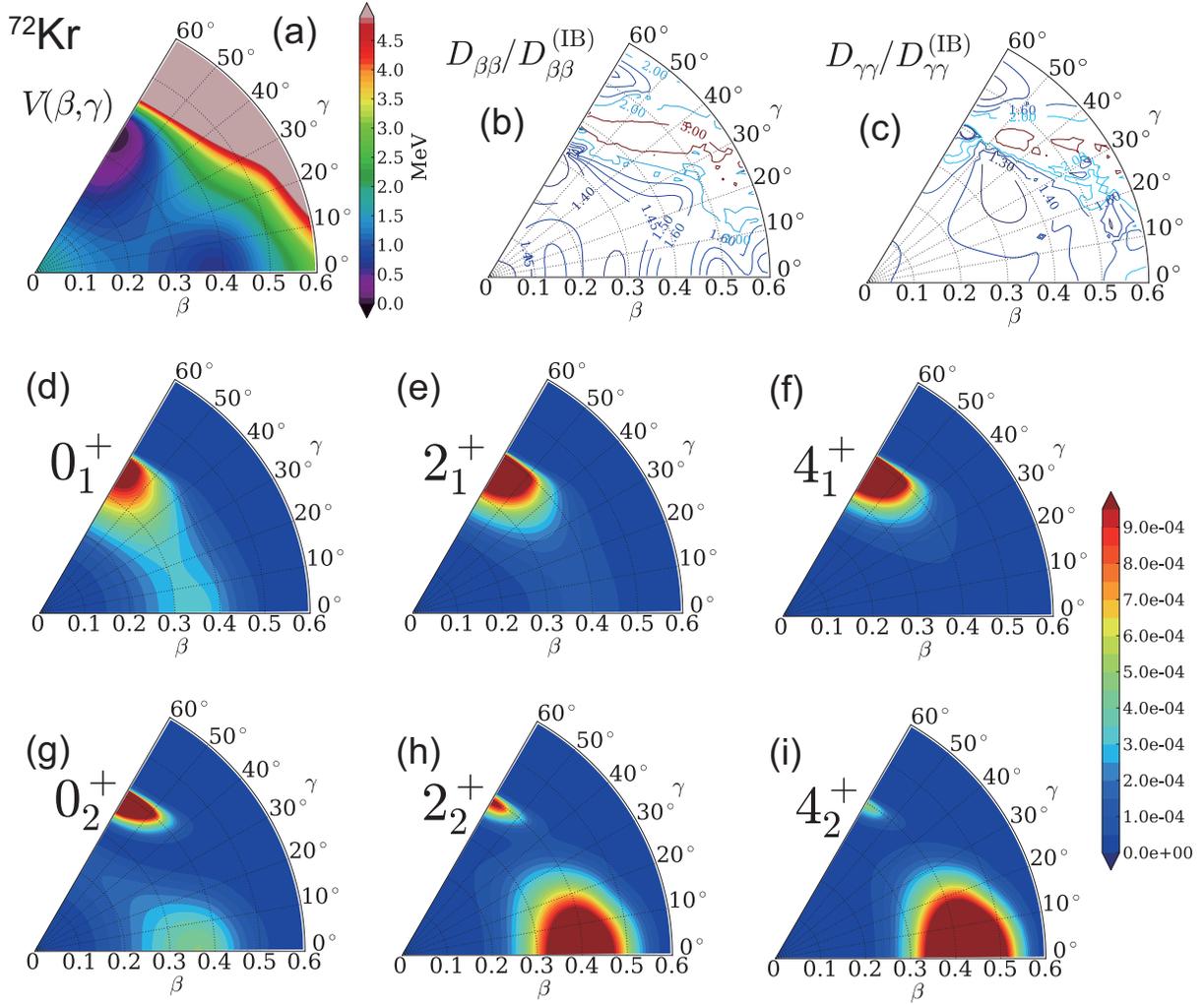}
\caption{ Application of the LQRPA method to the oblate-prolate shape coexistence/fluctuation phenomenon 
in $^{72}$Kr 
(from Ref.~\cite{sat11}). 
(a) Collective potential $V(\beta,\gamma)$, 
(b) Ratios of the collective inertial masses $D_{\beta\beta}(\bg)$ 
to the Inglis-Belyaev cranking masses. 
(c) Same as (b) but for $D_{\gamma\gamma}(\bg)$. 
Vibrational wave functions squared, $\sum_K \beta^4|\Phi_{\alpha IK}(\beta,\gamma)|^2$, 
for (d) the $0_1^+$ state,  
(e) the $2_1^+$ state, 
(f) the $4_1^+$ state, 
(g) the $0_2^+$ state, 
(h) the $2_2^+$ state, 
and (i) the $4_2^+$ state. 
For the $\beta^4$ factor, see the text. 
}
\label{fig:72Kr_wave}
\end{center}
\end{figure}

\clearpage
\begin{figure}[tbp]
\begin{center}
\includegraphics[width=\textwidth]{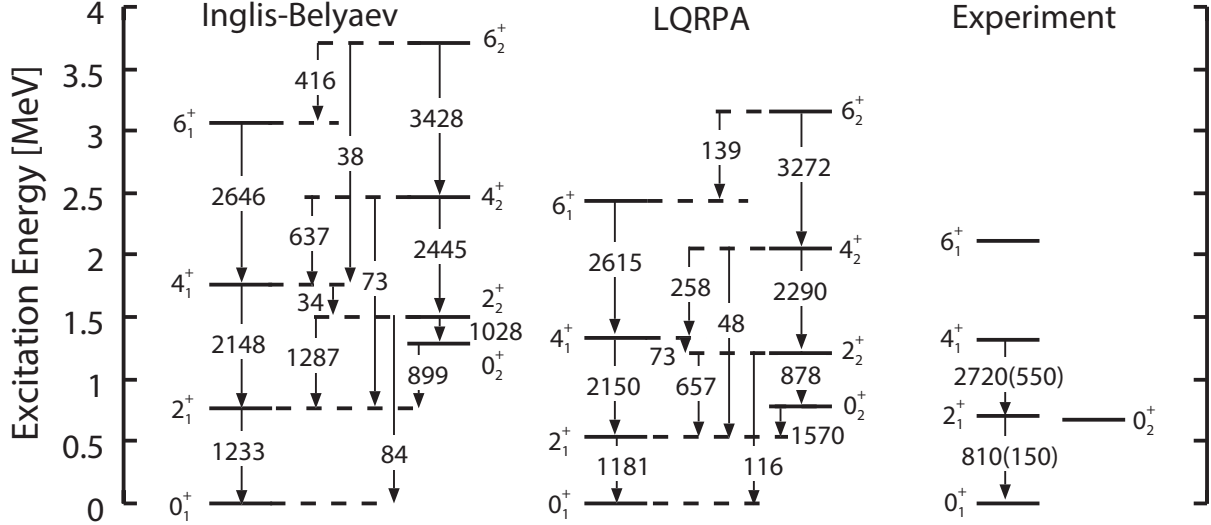}
\caption{
Excitation spectra and $B(E2)$ values calculated for $^{72}$Kr 
by means of the CHB+LQRPA method (denoted by LQRPA) \cite{sat11} 
and experimental data \cite{fis03, iwa14}.  
For comparison, the results calculated using the Inglis-Belyaev cranking mass 
are also shown. Only $B(E2)$fs larger than 1 Weisskopf unit are shown 
in units of $e^2 {\rm fm}^4$.
}
\label{fig:72Kr_spectra}
\end{center}
\end{figure}

The collective potential $V(\beta,\gamma)$ depicted in Fig.~1 exhibits two local minima. 
The oblate minimum is lower than the prolate minimum.  
This is expected from the deformed shell structure which gives rise to 
an oblate magic number at $Z=N=36$.   
This figure also shows that the valley runs in the triaxially deformed region and the barrier 
connecting the oblate and prolate minima is low. 
Accordingly, one may expect large-amplitude quantum shape fluctuations 
to occur along the triaxial valley. 
In fact, the vibrational wave function of the ground  $0^+_1$ state peaks around the oblate potential
minimum, but its tail extends to the prolate region. 
The vibrational wave function of the excited $0^+_2$ state
consists of two components: one is a sharp peak on the oblate side 
and the other is a component spreading around the prolate region somewhat broadly. 
It is interesting to notice that, as the angular momentum increases,  
the localization of the vibrational wave functions in the $(\beta, \gamma)$ deformation plane 
develops; 
namely, the rotational effect plays an important role 
for the emergence of the shape-coexistence character.  

We note that not only the vibrational inertial masses shown in Fig. 1 but also
the rotational inertial functions $(D_1, D_2$, and $D_3)$ and the pairing gaps  
significantly change as functions of $(\bg)$. 
Due to the time-odd contributions of the moving HFB self-consistent field, 
the collective inertial masses calculated with the LQRPA method are $20-50\%$ larger than 
those evaluated with the Inglis-Belyaev cranking formula. 
Their ratios also change as functions of $(\bg)$ 
\cite{sat11}. 
As a consequence, as shown in Fig.~2, 
the excitation spectrum calculated with the LQRPA masses is in much better 
agreement with experimental data than that with the Inglis-Belyaev cranking masses.
\\

\noindent
{\it Quantum shape transitions and fluctuations in $^{30,32,34}$Mg} 
\\

\begin{figure}[tbp]
\begin{center}
\includegraphics[width=\textwidth]{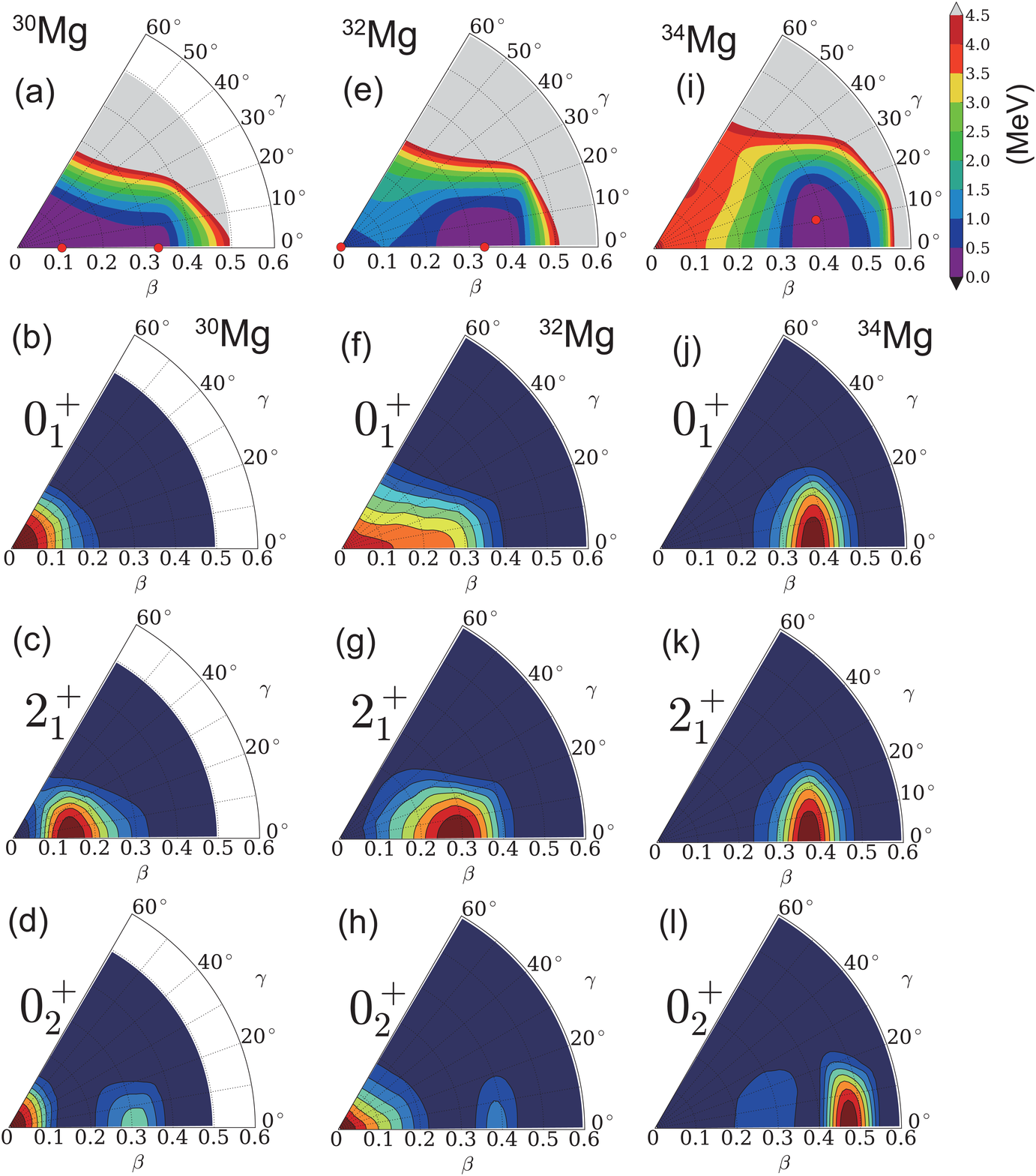}
\caption{
Application of the LQRPA method to the low-lying states in $^{30,32,34}$Mg 
(from Ref.~\cite{hin11b}). 
(a) Collective potential $V(\beta,\gamma)$ for $^{30}$Mg.  
The HFB equilibrium points are indicated by red circles. 
(b)-(d) Vibrational wave functions squared, $\sum_K |\Phi_{\alpha IK}(\beta,\gamma)|^2$, 
of the $0_1^+, 2_1^+$, and $0_2^+$ states in $^{30}$Mg. 
Contour lines are drawn at every tenth part of the maximum value. 
(e)-(h) and (i)-(l): Same set of figures but for $^{32}$Mg and $^{34}$Mg, respectively.   
}
\label{fig:Mg_wave}
\end{center}
\end{figure}

\begin{figure}[tbp]
\begin{center}
\includegraphics[width=0.4\textwidth]{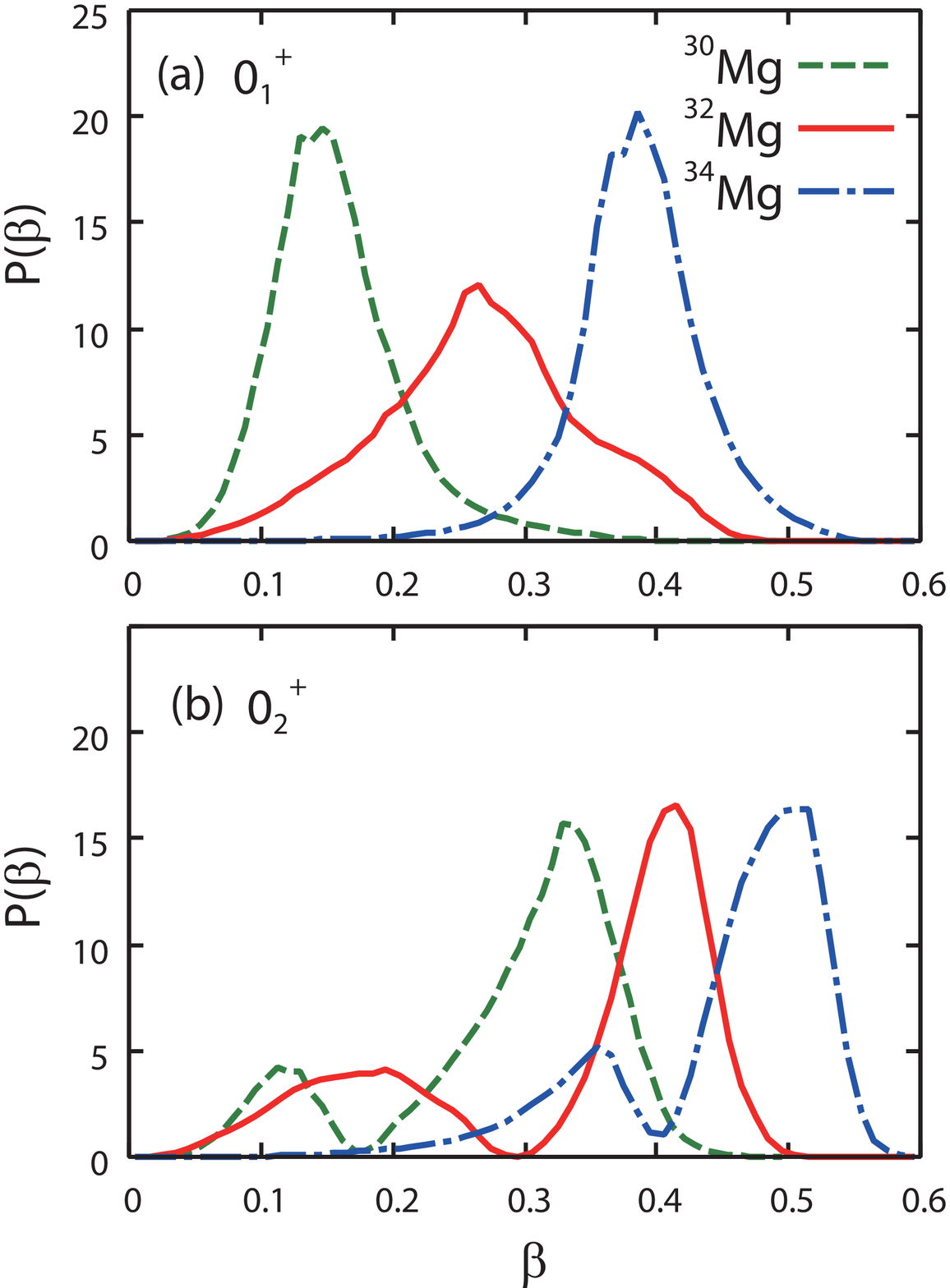}
\caption{
(a) Probability densities integrated over $\gamma$, 
$P(\beta)=\int d\gamma |\Phi_{\alpha I=0,K=0}(\beta,\gamma)|^2
 |G(\beta,\gamma|^{1/2}$, 
for the ground $0_1^+$ states in $^{30,32,34}$Mg 
plotted as functions of $\beta$. 
(b) Same as (a) but for the excited $0_2^+$ states.
}
\label{fig:Mg_PbI0}
\end{center}
\end{figure}

This is a new region of quantum shape transition currently under live discussions 
toward understanding the nature of the quadrupole deformation in these neutron-rich isotopes 
as well as  the mechanism of its emergence. 

Figure 3 shows the collective potentials $V(\beta,\gamma)$
and the vibrational wave functions squared, 
$\sum_K |\Phi_{\alpha IK}(\beta,\gamma)|^2$,
of the $0_1^+, 2_1^+$, and $0_2^+$ states in $^{30,32,34}$Mg.
It is clearly seen that prolate deformation grows with the neutron number. 
The deformed magic numbers, $Z=12$ of protons and 
$N=20,22,24$ at different values of $\beta$ of neutrons 
\cite{ham12} 
act cooperatively for the appearance of the prolate minima.
Interestingly, the vibrational wave functions of the $2_1^+$ state are noticeably different 
from those of the $0_1^+$ state in $^{30}$Mg and $^{32}$Mg, 
while they are similar in $^{34}$Mg. 

In Fig. 4, we display the probability density 
of finding a shape with a specific value of $\beta$. 
\begin{equation}
P(\beta)=\int d\gamma |\Phi_{\alpha I=0,K=0}(\beta,\gamma)|^2
 |G(\beta,\gamma)|^{1/2}. 
\end{equation}

Note that the volume element with  
$|G(\beta,\gamma)| =4\beta^8 W(\beta,\gamma) R(\beta,\gamma) \sin^2 3\gamma $  
is taken into account here.  
Let us first look at the ground $0_1^+$ states. 
The peak position moves toward a larger value of $\beta$ in going from 
$^{30}$Mg to $^{34}$Mg. 
The distribution for  $^{32}$Mg is much broader than those for $^{30}$Mg and $^{34}$Mg.
Next, let us look at the excited $0_2^+$ states.
In $^{30}$Mg, the bump at $\beta \simeq 0.1$ is much smaller than the major bump 
at $\beta \simeq 0.3$. In this sense, we can regard the $0_2^+$ state of $^{30}$Mg 
as a prolately deformed state. 
In the case of $^{32}$Mg, the probability density exhibits a very broad distribution 
extending from the spherical to deformed regions up to $\beta = 0.5$ with a prominent
peak at $\beta \simeq 0.4$ and a node at $\beta \simeq 0.3$. 

Thus, the shape coexistence picture that 
the deformed excited $0_2^+$ state coexists with the
spherical ground state approximately holds for $^{30}$Mg. 
On the other hand, large-amplitude quadrupole-shape fluctuations dominate in
both the ground and the excited $0^+$ states in $^{32}$Mg, 
in contrast to the interpretation of gdeformed ground and spherical
excited $0^+$ statesh 
\cite{wim10} 
based on a naive picture of crossing between the spherical and deformed configurations. 
To test these theoretical predictions, 
an experimental search for the distorted
rotational bands built on the excited $0_2^+$ states 
in $^{30}$Mg and $^{32}$Mg is strongly desired.
\\

\section{Some remarks on other approaches}

In Section 2, we reviewed the basics of a microscopic theory of LACM 
focusing on new developments achieved after 2000. 
In this section, we give short remarks on other approaches to  LACM.  
Typical approaches developed by 1980 are described in detail in the textbook of 
Ring and Schuck \cite{rin80},  
and achievements during 1980-2000 are well summarized in the review by 
Do Dang, Klein, and Walet \cite{dan00}. 
\\

\noindent
{\it Constrained HFB + adiabatic perturbation}\\
 
Historically, the Inglis-Belyaev cranking masses 
derived from the adiabatic perturbation theory 
\cite{rin80} 
have been widely used in conjunction with phenomenological 
mean-field models, e.g. for the study of fission dynamics 
\cite{bra72}.
In recent years, it has become possible to carry out such studies using 
self-consistent mean fields obtained by solving the constrained HFB equations 
\cite{bar11}. 
The Inglis-Belyaev cranking masses have also been used for 
low-frequency quadrupole collective dynamics 
\cite{lib99,yul99,pro04,del10}. 
At present, a systematic investigation on low-lying quadrupole spectra is underway 
using the 5D collective Hamiltonian with Inglis-Belyaev cranking masses 
and the relativistic (covariant) density functionals 
\cite{nik09, li09, li10a, li10b, li11,nik11, fu13}. 

A problem of the Inglis-Belyaev cranking formula is that time-odd mean-field effects 
are ignored and it underestimates the collective masses (inertial functions) 
\cite{dob95}. 
Moving mean fields possess time-odd components that change sign 
under time reversal operation,  
but the cranking formula ignores their effects on the collective masses.    
By taking into account such time-odd corrections to the cranking masses,    
one can better reproduce low-lying spectra 
\cite{hin12}. 
\\

\noindent
{\it Adiabatic TDHF theory}\\

Attempts to self-consistently derive the collective Hamiltonian  
using adiabatic approximation to time evolution of mean fields 
started in 1960's 
\cite{bar65, bel65}. 
In these pioneer works, the collective quadrupole coordinates $(\beta,\gamma)$ 
were defined in terms of expectation values of the quadrupole operators and 
the 5D collective Hamiltonian was derived 
using the P+Q force model 
\cite{bes69}.  
During 1970's this approach was generalized to be applicable to any effective interaction. 
This advanced approach is called adiabatic TDHF (ATDHF) 
\cite{bar78,bri76,goe78}. 

In the ATDHF theory, the density matrix $\rho(t)$ is written in the following form and 
expanded as a power series with respect to $\chi(t)$. 
\begin{eqnarray}
\rho(t)&=&e^{i\chi(t)}\rho_0(t)e^{-i\chi(t)}\\
          &=&\rho_0(t)+i[\chi(t),\rho_0(t)]-\frac{1}{2}[\chi(t), [\chi(t),\rho_0(t)]] + ... 
\end{eqnarray}
Baranger and V\'en\'eroni \cite{bar78} 
suggested a possibility to introduce collective coordinates 
as parameters describing the time evolution of the density matrix $\rho_0(t)$ 
and discussed an iterative procedures to solve the ATDHF equations.  
But this idea has not been realized until now. 
We note that the ATDHF does not reduce to the RPA in the small-amplitude limit 
if a few collective coordinates are introduced by hand. 
In fact it gives a collective mass different from the RPA 
\cite{gia80}. 

The ATDHF theory developed by Villars 
\cite{vil77} 
aims at self-consistent determination of 
the optimum collective coordinates on the basis of the time-dependent variational principle.  
This approach, however, encountered a difficulty that we could not get 
unique solutions of its basic equations determining the collective path. 
This problem was later solved by treating the second-order terms of the 
momentum expansion in a self-consistent manner 
\cite{kle91a, muk81}. 
It was shown that, when the number of collective coordinate is only one,  
a collective path maximally decoupled from non-collective degrees of freedom 
runs along a valley in the multi-dimensional potential-energy surface 
associated with the TDHF states.   

To describe low-frequency collective motions, 
it is necessary to take into account the pairing correlations. 
In other words, we need to develop the adiabatic TDHFB (ATDHFB) theory. 
This is one of the reasons why applications of the ATDHF have been restricted 
to collective phenomena where pairing correlations play minor roles such as 
low-energy collisions between spherical closed-shell nuclei 
\cite{goe83} 
and giant resonances 
\cite{gia80}. 
When large-amplitude shape fluctuations take place, 
single-particle level crossings often occur. 
To follow the adiabatic configuration across the level crossing points, 
the pairing correlation plays an essential role. 
Thus, an extension to ATDHFB is indispensable for 
the description of low-frequency collective excitations.  

In the past, 
Dobaczewski and Skalski \cite{dob81} 
tried to develop the ATDHFB theory 
assuming the axially symmetric quadrupole deformation parameter $\beta$ as the collective coordinate. 
Quite recently, 
Li {\it et al.} \cite{li12a} 
tried to derive the 5D quadrupole collective Hamiltonian 
on the basis of the ATDHFB.  
The extension of ATDHF to ATDHFB is not straightforward, however. 
This is because, as discussed in Section 2, 
we need to decouple the number-fluctuation degrees of freedom from the LACM of interest 
and respect the gauge invariance with respect the pairing rotational angles. 
\\

\noindent
{\it Boson expansion method}\\

Boson expansion method is well known as a useful microscopic method 
of describing anharmonic (non-linear) vibrations going 
beyond the harmonic approximation of the QRPA.   
In this approach, we first construct a collective subspace 
spanned by many-phonon states of vibrational quanta (determined by the QRPA)  
in the huge-dimensional shell-model space, and then map these many-phonon states 
one-to-one to many-boson states in an ideal boson space. 
Anharmonic effects neglected in the QRPA are treated as higher-order terms 
in the power-series expansion with respect to the boson creation and annihilation operators. 
Starting from the QRPA about a spherical shape, one can thus derive the 5D quadrupole collective Hamiltonian 
in a fully quantum mechanical manner.  
The boson expansion method has been successfully applied to low-energy quadrupole excitation spectra 
in a wide range of nuclei including those lying in regions of quantum phase transitions 
from spherical to deformed 
\cite{sak88,kle91b}.  

In the time-dependent mean-field picture, 
state vectors in the boson expansion method are written 
in terms of the creation and annihilation operators
$(\Gamma_i^\dag, \Gamma_i)$ of the QRPA eigen-modes, 
or, equivalently, in terms of the collective coordinate and momentum operators 
$({\hat Q}_i, {\hat P}_i)$, 
\begin{eqnarray}
|\phi(t)\rangle &=& 
{\rm exp} \left[\sum_i \left( \eta_i(t) \Gamma_i - \eta_i^*(t) \Gamma_i^\dag \right)\right]
  |\phi_0\rangle \\
&=& {\rm exp}\left[ \sum_i \left( p_i(t) {\hat Q}_i - q_i(t) {\hat P}_i \right)\right] 
 |\phi_0\rangle.
\end{eqnarray}
With increasing amplitudes of the quadrupole shape vibration $|\eta_i(t)|$
(values of the collective coordinate $q_i(t)$), anharmonic (non-linear) effects become stronger. 
Strong non-linear effects may eventually change even the microscopic structure of the collective operators 
$({\hat Q}_i, {\hat P}_i)$ determined by the QRPA. 
In such situations, it is desirable to construct a theory that allows 
variations of microscopic structure of collective operators as functions of $q_i(t)$. 
It may be said that the SCC method has accomplished this task.  
\\

\noindent
{\it Generator coordinate method}\\

The  generator coordinate method (GCM) has been used 
for a wide variety of nuclear collective phenomena 
\cite{rei87, egi04, ben08a}. 
Using the angular-momentum projector $\hat{P}_{IMK}$ 
and the neutron(proton)-number projector 
$\hat{P}_N$ ($\hat{P}_Z$),  we write the state vector as a superpositions of the projected 
mean-field states with different deformation parameters $(\beta,\gamma)$,  
\begin{equation}
\ket{\Psi^i_{NZIM}} =  \int d\beta d\gamma \sum_K f^i_{NZIK} (\beta,\gamma) 
\hat{P}_N \hat{P}_Z \hat{P}_{IMK}  \ket{\phi(\beta,\gamma)}.  
\label{eq:GCM}
\end{equation}
Because the projection operators contain integrations, 
it has been a difficult task to carry out such high-dimensional numerical integrations 
in solving the Hill-Wheeler equation for the states $\ket{\phi(\beta,\gamma)}$
obtained by the constrained HFB method.    
In recent years, however, remarkable progress has been taking place, 
which makes it possible to carry out such large-scale numerical computations  
\cite{ben08b, rod10, yao10, yao11, yao14, rod14}. 
The HFB calculations using the density dependent effective interactions 
are better founded by density functional theory (DFT). 
Accordingly, the modern GCM calculation is referred to as `multi-reference DFT' 
\cite{ben08b}.  


It is well known that one can derive a collective Schr\"odinger equation 
by making a gaussian overlap approximation (GOA) to the Hill-Wheeler equation 
\cite{gri57,oni75,roh12}. 
There is no guarantee, however, that dynamical effects associated 
with time-odd components of moving mean field are sufficiently taken into account 
in the collective masses (inertial functions) obtained through this procedure. 
In the case of center of mass motion, 
we need to use complex generator coordinates to obtain the correct mass, 
implying that collective momenta conjugate to collective coordinates
should also be treated as generator coordinates 
\cite{rin80, pei62}. 

A fundamental question is how to choose the optimal generator coordinates. 
With the variational principle,  
Holzwarth and Yukawa \cite{hol74} 
proved that 
the mean-field states parametrized by a single optimal generator coordinate 
run along a valley of the collective potential energy surface.  
This line of investigation was further developed 
\cite{rei79} 
and greatly stimulated 
the challenge toward constructing a microscopic theory of LACM. 
In this connection, 
we note that conventional GCM calculations parametrized by a few real generator coordinates 
do not reduce to the (Q)RPA in the small-amplitude limit, 
differently from the case that all two-quasiparticle (particle-hole) degrees of freedom are treated as 
complex generator coordinates 
\cite{jan64}. 

It is very important to distinguish the 5D collective Hamiltonian obtained 
by making use of the GOA to the  GCM 
from that derived in Section 3 on the basis of the ASCC method. 
In the latter, the canonical conjugate pairs of collective coordinate and momentum 
are self-consistently derived on the basis of the time-dependent variational principle. 
The canonical formulation  enables us to adopt the standard canonical quantization procedure.  
Furthermore, effects of the time-odd components of the moving mean field 
are automatically taken into account in the collective masses (inertial functions). 
In view of the above points, it is highly desirable to carry out  
a systematic comparison of collective inertial masses evaluated by different approximations 
including the ASCC, the ATDHFB, the GCM+GOA, and the adiabatic cranking methods 
for a better understanding of their physical implications.   
In this connection, it is interesting to notice 
that the results of the recent GCM calculation for $^{76}$Kr 
\cite{yao14}, 
using the particle-number and angular-momentum projected basis~(\ref{eq:GCM}), 
are rather similar to those obtained by use of the 5D collective Hamiltonian 
with the Inglis-Belyaev cranking masses,    
except for an overall overestimation of the excitation energies by about 20$\%$. 

\section{Challenges for future}

As reviewed by Hyde and Wood 
\cite{hey11},  
nature of low-lying excited $0^+$ states, 
systematically found in recent experiments 
as well as those known from old days,
is not well understood. 
It is thus quite challenging to apply, 
in a systematic ways,  
the 5D collective Hamiltonian approach to all of 
these data, from light to heavy and 
from stable to unstable nuclei, 
and explore the limit of its applicability. 
Recalling that the importance of the couplings 
between the quadrupole and pairing vibrations has been pointed out 
\cite{iwa76,wee81,tak86}, 
one of the basic questions is `under what situations 
we need to extend the 5D collective Hamiltonian to 7D 
by explicitly treating the proton and neutron pairing gaps 
as dynamical variables.'    

Another interesting subject is to extend the collective Hamiltonian approach 
to a variety of collective phenomena, 
for example, those observed in rapidly rotating nuclei, 
heavy and super heavy nuclei, neutron-rich unstable nuclei, 
by taking into account the effects of rapid rotation and/or continuum, 
from the beginning in the single-particle (HFB) Hamiltonian. 
Macroscopic quantum tunnelings through self-consistently generated barriers,  
such as spontaneous fissions and deep sub-barrier fusions, are, needless to say, 
longstanding yet modern, fundamental subjects of nuclear structure physics.     

It is a great challenge to develop the CHFB + LQRPA approach 
on the basis of the time-dependent DFT.   
To efficiently solve the large-dimensional LQRPA equations containing 
density-dependent terms,    
the finite-amplitude method recently developed in 
Refs.~\cite{nak07,avo11,avo13,hin13} 
may be utilized. 




\begin{thebibliography}{999}
\bibitem{boh75} 
 A.~Bohr and B.~R.~Mottelson,@{\it Nuclear Structure}, Vol.~II, 
 (W.~A.~Benjamin Inc., 1975; World Scientific 1998).  

\bibitem{abe90}
 S. {\AA}berg, H. Flocard, and W. Nazarewicz,  
 Annu. Rev. Nucl. Part. Sci. {\bf 40} (1990), 439.
 
\bibitem{ben03} 
 M. Bender, P.-H. Heenen, and P.-G. Reinhard, 
 Rev. Mod. Phys. {\bf 75} (2003), 121. 

\bibitem{row10} 
 D. J. Rowe and J.L. Wood, 
 {\it Fundamentals of Nuclear Models, Foundational Models},  (World Scientific, 2010).    

\bibitem{hey11}
 K. Heyde and J.L. Wood, Rev. Mod. Phys. {\bf 83} (2011) 1467.

\bibitem{and58}
 P. W. Anderson,  Phys. Rev. {\bf 112} (1958), 1900; Phys. Rev. {\bf 130} (1963), 439. 

\bibitem{nam60} 
 Y. Nambu, Phys. Rev. {\bf 117} (1960), 648.   

\bibitem{bri05}
 D. M. Brink and R. A. Broglia,  {\it Nuclear Superfluidity, Pairing in Finite Systems} 
  (Cambridge University Press, 2005). 

\bibitem{fra01}
 S. Frauendorf, Rev. Mod. Phys. {\bf 73} (2001), 463.

\bibitem{boh76} 
 A. Bohr, Rev. Mod. Phys. {\bf 48} (1976), 365.          

\bibitem{mot76}  
 B. Mottelson, Rev. Mod. Phys. {\bf 48} (1976), 375.  
 
\bibitem{pro09} 
 L. Pr{\'o}chniak and S. G. Rohozi{\'n}ski, 
 J. Phys. G. Nucl. Part. Phys. {\bf 36} (2009), 123101.

\bibitem{mat85b} 
 M. Matsuo and K. Matsuyanagi, 
 Prog. Theor. Phys. {\bf 74} (1985), 1227; 
 Prog. Theor. Phys. {\bf 76} (1986), 93; 
 Prog. Theor. Phys. {\bf 78} (1987), 591. 

\bibitem{iwa76} 
S. Iwasaki, T. Marumori, F. Sakata and K. Takada,  
Prog. Theor. Phys. {\bf 56} (1976), 1140. 

\bibitem{wee81}  
 K. J. Weeks, T. Tamura, T. Udagawa, F.J.W. Hahne, 
 Phys. Rev. C {\bf 24} (1981), 703.

\bibitem{tak86}  
K. Takada and S. Tazaki, Nucl. Phys. A {\bf 448} (1986), 56.

\bibitem{neg82} 
 J. W. Negele, Rev. Mod. Phys. {\bf 54} (1982), 913.

\bibitem{abe83} 
 A. Abe and T. Suzuki (ed.) ,
 {\it Microscopic Theories of Nuclear Collective Motions},  
 Prog. Theor. Phys. Suppl. Nos. {\bf 74} \&{\bf 75} (1983).

\bibitem{yam87} 
 M. Yamamura and A. Kuriyama, Prog. Theor. Phys. Suppl. No. {\bf 93} (1987). 

\bibitem{kur01}
 A.~Kuriyama, K.~Matsuyanagi, F.~Sakata, K.~Takada, and M.~Yamamura (ed.), 
 {\it Selected Topics in the Boson Mapping and Time-Dependent Hartree--Fock Methods}, 
 Prog. Theor. Phys. Suppl. No. {\bf 141} (2001), 1

\bibitem{nak99} 
 T. Nakatsukasa, N.R. Walet and G. Do Dang, Phys. Rev. C {\bf 61}  (1999), 014302.    

\bibitem{rin80} 
 P. Ring and P. Schuck, {\it The Nuclear Many-Body Problem}  (Springer-Verlag, 1980).

\bibitem{bla86}
 J.-P. Blaizot and G. Ripka, 
  {\it Quantum Theory of Finite Systems} (The MIT press, 1986). 
 
\bibitem{mat10} 
 K. Matsuyanagi, M. Matsuo, T. Nakatsukasa, N. Hinohara, and K. Sato, 
 J. Phys. G: Nucl. Part. Phys. {\bf 37} (2010), 064018. 

\bibitem{nak12} 
 T. Nakatsukasa, Prog. Theor. Exp. Phys.  01A207 (2012). 
 
\bibitem{mat13} 
 K. Matsuyanagi, N. Hinohara, and K. Sato, 
 in ``{\it Fifty Years of Nuclear BCS: Pairing in Finite Systems}"
 (World Scientific 2013).          

\bibitem{row76} 
 D. J. Rowe and R. Bassermann, Canad. J. Phys. {\bf 54} (1976), 1941.

\bibitem{mar77} 
 T. Marumori, Prog. Theor. Phys. {\bf 57} (1977), 112.     

\bibitem{mar80} 
 T. Marumori, T. Maskawa, F. Sakata, and  A. Kuriyama, 
 Prog. Theor. Phys. {\bf 64} (1980), 1294.

\bibitem{row82} 
 D. J. Rowe, Nucl. Phys. {\bf 391} (1982), 307.

\bibitem{mat86} 
 M. Matsuo, Prog. Theor. Phys. {\bf 76} (1986), 372. 

\bibitem{mat85a} 
 M. Matsuo and K. Matsuyanagi, 
 Prog. Theor. Phys. {\bf 74} (1985), 288. 
 
\bibitem{mat85c} 
 M. Matsuo, Y. R. Shimizu, and K. Matsuyanagi, 
 {\it Proceedings of The Niels Bohr Centennial Conf.  on Nuclear Structure}, 
 ed. R.~Broglia, G.~Hagemann and  B.~Herskind (North-Holland, 1985), p.~161.

\bibitem{yam93}
 K. Yamada, Prog. Prog. Theor. Phys. {\bf 89} (1993), 995.

\bibitem{shi01} 
 Y. R. Shimizu and K. Matsuyanagi,  
 Prog. Theor. Phys. Suppl. No. {\bf 141} (2001), 285. 

\bibitem{mat00} 
 M. Matsuo, T. Nakatsukasa, and K. Matsuyanagi, 
 Prog. Theor. Phys. {\bf 103} (2000), 959.

\bibitem{kle91a} 
 A. Klein, N.R. Walet, and G. Do Dang, Ann. of Phys.  {\bf 208} (1991), 90. 

\bibitem{alm04a} 
 D. Almehed and N. R. Walet, Phys. Rev. C {\bf 69} (2004), 024302.  

\bibitem{bar90}
 F. Barranco, G.F. Bertsch, R.A. Broglia, and E. Vigezzi, 
 Nucl. Phys. A {\bf 512} (1990), 253. 

\bibitem{hin07} 
 N. Hinohara, T. Nakatsukasa, M. Matsuo, and K. Matsuyanagi, 
 Prog. Theor. Phys. {\bf 117} (2007), 451. 

\bibitem{hin10}
 N. Hinohara, K. Sato, T. Nakatsukasa, M. Matsuo, and K. Matsuyanagi, 
 Phys. Rev. {\bf C 82} (2010), 064313. 

\bibitem{sat11}  
 K. Sato and N. Hinohara, Nucl. Phys. A {\bf 849} (2011), 53.

\bibitem{sat12}  
 K. Sato, N. Hinohara, K. Yoshida, T. Nakatsukasa, M. Matsuo, and K. Matsuyanagi,  
 Phys. Rev. {\bf C 86} (2012), 024316.

\bibitem{bar65}
 M. Baranger and K. Kumar, Nucl. Phys. {\bf 62} (1965) 113; 
 Nucl. Phys. A {\bf 110} (1968), 529; Nucl. Phys. A {\bf 122} (1968), 241;  
 Nucl. Phys. A {\bf 122} (1968), 273. 

\bibitem{tho62}
 D. J. Thouless and J. G. Valatin, Nucl. Phys. {\bf 31} (1962), 211. 
 
\bibitem{eis87} 
 J.M. Eisenberg and W. Greiner, {\it Nuclear Theory}, Vol. 1, 3rd ed. 
 (North Holland, 1987).

\bibitem{kum67}
 K. Kumar and M. Baranger, Nucl. Phys. {\bf 92} (1967) 608. 

\bibitem{bes69}
 D. R. Bes and R. A. Sorensen, 
{\it Advances in Nuclear Physics} vol. 2,  (Prenum Press, 1969), p.~129. 

\bibitem{hin11a}
 N. Hinohara, K. Sato, T. Nakatsukasa, M. Matsuo, and K. Matsuyanagi, 
 Phys. Rev. {\bf C 84} (2011), 061302(R).

\bibitem{yos11}  
K. Yoshida and N. Hinohara, Phys. Rev. C {\bf 83} (2011), 061302(R).   


\bibitem{hin09} 
 N. Hinohara, T. Nakatsukasa, M. Matsuo, and K. Matsuyanagi,   
 Phys. Rev. C {\bf 80},    (2009), 014305.

\bibitem{hin11b}
N. Hinohara and Y. Kanada-En'yo, Phys. Rev. C {\bf 83} (2010), 014321.  

\bibitem{hin12} 
 N. Hinohara, Z. P. Li, T. Nakatsukasa, T. Nik\v{s}i\'{c}, and D. Vretenar, 
 Phys. Rev. C {\bf 85} (2012), 024323.  
 
\bibitem{fis03}
 S. M. Fischer {\it et al.}, Phys. Rev. C {\bf 67} (2003), 064318.

\bibitem{iwa14} 
H. Iwasaki {\it et al.},  Phys. Rev. Lett. {\bf 112} (2014), 142502. 

\bibitem{ham12}
 I. Hamamoto,  Phys. Rev. C {\bf 85} (2012), 064329.  

\bibitem{wim10} 
 K. Wimmer {\it et al.},    Phys. Rev. Lett. {\bf 105} (2010), 252501.

\bibitem{dan00} 
 G. Do Dang, A. Klein, and N.R. Walet, Phys. Rep.  {\bf 335} (2000), 93. 

\bibitem{bra72}
 M. Brack, Jens Damgaard, A.S. Jensen, H.C. Pauli,  V.M. Strutinsky, 
 and C.Y. Wong, Rev. Mod. Phys. {\bf 44} (1972), 320. 

\bibitem{bar11} 
 A. Baran, J. A. Sheikh, J. Dobaczewski, W. Nazarewicz, and A. Staszczak, 
 Phys. Rev. C {\bf 84}, (2011), 054321. 

\bibitem{lib99} 
 J. Libert, M. Girod, and J.-P. Delaroche,  Phys. Rev. C {\bf 60} (1999), 054301. 

\bibitem{yul99} 
 E.Kh. Yuldashbaeva, J. Libert, P. Quentin, and M. Girod,  
 Phys. Lett. B {\bf 461} (1999), 1. 

\bibitem{pro04} 
 L. Pr{\'o}chniak, P. Quentin, D. Samsoen,  and J. Libert, 
 Nucl. Phys. A. {\bf 730} (2004), 59.

\bibitem{del10}
 J.-P. Delaroche, M. Girod, J. Libert, H. Goutte, S. Hilaire, S. P\'{e}ru, N. Pillet, 
 and G. F. Bertsch, Phys. Rev. C {\bf 81} (2010), 014303.  

\bibitem{nik09} 
 T. Nik\v{s}i\'{c}, Z. P. Li, D. Vretenar, L. Pr{\'o}chniak, J. Meng, and P. Ring,  
 Phys. Rev. C {\bf 79} (2009), 034303.

\bibitem{li09}
 Z. P. Li, T. Nik\v{s}i\'{c}, D. Vretenar, J. Meng, G. A. Lalazissis, and P. Ring, @
 Phys. Rev. C {\bf 79} (2009), 054301. 

\bibitem{li10a}
 Z. P. Li, T. Nik\v{s}i\'{c}, D. Vretenar, and J. Meng, 
 Phys. Rev. C {\bf 81} (2010), 034316.  

\bibitem{li10b}
 Z. P. Li, T. Nik\v{s}i\'{c}, D. Vretenar, P. Ring, and J. Meng, 
 Phys. Rev. C {\bf 81} (2010), 064321. 

\bibitem{li11} 
 Z. P. Li, J. M. Yao, D. Vretenar, T. Nik\v{s}i\'{c}, H. Chen, and J. Meng, 
 Phys. Rev. C {\bf 84} (2011), 054304.  

\bibitem{nik11} 
 T. Nik\v{s}i\'{c}, D. Vretenar, and P. Ring,  Prog. Part. Nucl. Phys. {\bf 66} (2011), 519.

\bibitem{fu13}
Y. Fu, H. Mei, J. Xiang, Z.P. Li, J.M.Yao, and J. Meng, 
Phys. Rev. C {\bf 87} (2013), 054305. 

\bibitem{dob95}
 J. Dobaczewski and J. Dudek, Phys. Rev. C {\bf 52} (1995), 1827. 
 
\bibitem{bel65}
 S. T. Belyaev, Nucl. Phys. {\bf 64} (1965) 17. 

\bibitem{bar78}
 M. Baranger and M. V\'en\'eroni, Ann. of Phys. {\bf 114} (1978), 123.

\bibitem{bri76} 
 D. M. Brink, M. J. Giannoni, and M. Veneroni, Nucl. Phys. A {\bf 258} (1976), 237.

\bibitem{goe78}
 K. Goeke and P.-G. Reinhard, Ann. of Phys. {\bf 112} (1978), 328.

\bibitem{gia80}
 M. J. Giannoni and P. Quentin, 
 Phys. Rev. C {\bf 21} (1980), 2060; Phys. Rev. C {\bf 21} (1980), 2076.

\bibitem{vil77} 
 F. Villars, Nucl. Phys. A {\bf 285} (1977), 269. 

\bibitem{muk81} 
 A. K. Mukherjee and M. K. Pal, Nucl. Phys. A {\bf 373} (1982), 289.

\bibitem{goe83}
 K. Goeke, R. Y. Cusson, F. Grummer, P.-G. Reinhard, and H. Reinhardt, 
 Prog. Theor. Phys. Suppl.  No. {\bf 74 \& 75} (1983), 33. 

\bibitem{dob81} 
 J. Dobaczewski and J. Skalski, Nucl. Phys.  A {\bf 369} (1981), 123.

\bibitem{li12a}
 Z. P. Li, T. Nik\v{s}i\'{c}, P. Ring, D. Vretenar, J. M. Yao, and J. Meng, 
 Phys. Rev. C {\bf 86} (2012), 034334. 

\bibitem{sak88} 
 H. Sakamoto and T. Kishimoto, Nucl. Phys. A {\bf 486} (1988), 1. ;   
 Nucl. Phys. A {\bf 528} (1991), 73. 


\bibitem{kle91b}
 A. Klein and E. R. Marshalek, Rev. Mod. Phys. {\bf 63} (1991), 375. 

\bibitem{rei87} 
 P.-G.Reinhard and K. Goeke, Rep. Prog. Phys. {\bf 50} (1987), 1.   

\bibitem{egi04} 
 J. L. Egido and L. M. Robledo, in {\it Extended Density Functionals
 in Nuclear Physics}, edited by G. A. Lalazissis, P. Ring, and
 D. Vretenar, Lecture Notes in Physics, Vol. {\bf 641} (Springer, 2004), p. 269.

\bibitem{ben08a} 
 M. Bender, Eur. Phys. J. Spec. Top. {\bf 156} (2008), 217.

\bibitem{ben08b}
 M. Bender and P.-H. Heenen, Phys. Rev. C {\bf 78} (2008), 024309.  

\bibitem{rod10} 
 T. R. Rodr{\'i}guez and J. L. Egido, Phys. Rev. C {\bf 81} (2010), 064323.  

\bibitem{yao10} 
 J. M. Yao, J. Meng, P. Ring, and D. Vretenar,  
 Phys. Rev. C {\bf 81} (2010), 044311.  

\bibitem{yao11} 
 J. M. Yao, H. Mei, H. Chen, J. Meng, P. Ring, and D. Vretenar, 
 Phys. Rev. C {\bf 83} (2011), 014308.  

\bibitem{yao14} 
 J. M. Yao, K. Hagino, Z. P. Li, J. Meng, and P. Ring, 
 Phys. Rev. C {\bf 89} (2014), 054306.  

\bibitem{rod14} 
 T. R. Rodr{\'i}guez Phys. Rev. C {\bf 90} (2014), 034306.  

\bibitem{gri57}
 J.J. Griffin and J.A. Wheeler, Phys. Rev. {\bf 108} (1957), 311.

\bibitem{oni75} 
 N. Onishi and T. Une, Prog. Theor. Phys. {\bf 53} (1975), 504.

\bibitem{roh12} 
 S. G. Rohozi{\'n}ski, J. Phys. G. Nucl. Part. Phys. {\bf 39} (2012), 095104.

\bibitem{pei62} 
 R. E. Peierls and D. J. Thouless , Nucl. Phys. {\bf 38} (1962), 154.

\bibitem{hol74} 
 G. Holzwarth and T. Yukawa, Nucl. Phys. A {\bf 219} (1974), 125. 

\bibitem{rei79} 
 P.-G. Reinhard and K. Goeke, Phys. Rev. C {\bf 20} (1979), 1546.   

\bibitem{jan64} 
 B. Jancovici and D.H. Schiff, Nucl. Phys. {\bf 58} (1964), 678. 
 
\bibitem{nak07} 
 T. Nakatsukasa, T. Inakura, and K. Yabana, Phys. Rev. C {\bf 76} (2007). 024318

\bibitem{avo11}
 P. Avogadro and T. Nakatsukasa, Phys. Rev. C {\bf 84}(2011), 014314.

\bibitem{avo13}
 P. Avogadro and T. Nakatsukasa, Phys. Rev. C {\bf 87}(2013), 014331.

\bibitem{hin13}
N. Hinohara, M. Kortelainen, and W. Nazarewicz, 
Phys. Rev. C {\bf 87} (2013), 064309. 


\end{thebibliography}
\end{document}